\documentclass[10pt, a4paper]{article}
\usepackage[a4paper]{geometry}
\usepackage{url}
\usepackage[pdftex]{graphicx}
\usepackage[linesnumbered,ruled]{algorithm2e}
\usepackage{amsmath}

\begin{document}

\author{
  Dimitrios Vasilas \\[-0.2ex]
  {\small Scality, Sorbonne Universit\'e - LIP6 \& Inria, France}
  \and
  Marc Shapiro \\[-0.2ex]
  {\small Sorbonne Universit\'e - LIP6 \& Inria, France}
  \and
  Bradley King\\[-0.2ex]
  {\small Scality, France}
}
\date{}

\title{A Modular Design for Geo-Distributed Querying
 \\ \vskip 0.5em \large{Work in Progress Report}}

\maketitle

\begin{abstract}

Most distributed storage systems provide limited abilities for querying data by
attributes other than their primary keys.
Supporting efficient search on secondary attributes is challenging as
applications pose varying requirements to query processing systems,
and no single system design can be suitable for all needs.
In this paper, we show how to overcome these challenges in order to
extend distributed data stores to support queries on secondary
attributes.
We propose a modular architecture that is flexible and allows query processing
systems to make trade-offs according to different use case requirements.
We describe adaptive mechanisms that make use of this flexibility to enable
query processing systems to dynamically adjust to query and write operation
workloads.

\end{abstract}


\section{Introduction}
Large-scale distributed storage systems, such as BigTable, Dynamo, Cassandra and
HBase among others, are an essential component of cloud computing applications.
While these storage systems offer significant performance and scalability
advantages to applications, they achieve these properties by exposing a simple
API that allows objects to be retrieved only by using the key under which they
were inserted.
However, applications often require the ability to access data by performing
queries on secondary attributes.
For example, a media application that manages video and image files, can benefit
from the ability to tag and retrieve objects using attributes describing the
content type, resolution format, author, copyright information, and other
descriptive attributes.
The restrictive key-based API makes it difficult and inefficient to implement
applications that require this functionality.

A common approach to address this problem is to maintain indexes in order to
achieve high-performance queries on secondary attributes.
Indexing has been studied extensively in systems offering strong consistency
guarantees, especially in the context of traditional database systems.
However, modern NoSQL-style data stores use geo-replication across several data
centres (DCs) and implement weak consistency models in which client operations
are served by  accessing the local replica without synchronising with other DCs,
in order to avoid wide-area network latency and remain available under
partition.

Indexing in this context poses several unique challenges.
Guided by the storage systems' decision to favour availability and low latency
over consistency, secondary indexes also need to use replication and be able to
ingest updates and process queries concurrently in different replicas without
requiring synchronisation.
Weakly consistent indexes must ensure that index replicas converge to the same
state under concurrent conflicting updates, and that the converged state is
consistent with the base data.
Moreover, since these storage systems are designed to achieve low-latency,
index maintenance should not penalise write latency.

Secondary indexing is not suitable for all cases, since maintaining
indexes incurs significant storage and maintenance costs.
In cases where queries on secondary attributes are not frequent and search
latency in not critical, other approaches such as scanning the underlying data
store can be acceptable.

No single query processing system design can be appropriate for all uses.
Design decisions cannot be made in isolation from data distribution and
replication schemes \cite{Dsilva2017SecondaryIT}.
At the same time, querying systems need to optimise different metrics, such as
query response latency or freshness, or provide specific types of queries
depending on the needs of each application.
Existing approaches \cite{196190, Escriva:2012:HDS:2342356.2342360,
Tai:2016:RSH:3026959.3026991} are implemented by making design decisions
targeting specific use case requirements and system characteristics.
A promising direction can be a flexible approach that enables query processing
systems to make different trade-offs and adjust to the needs of each use case.

In this paper, we propose a modular architecture for extending geo-distributed
data stores with the functionality to query data using secondary attributes.
Instead of making design decisions based on a specific system model or
application use case, here we propose a flexible architecture where querying
systems are built by interconnecting fine-grained software components.
Different querying system configurations can be built using the same building
blocks, each making different trade-offs and targeting specific requirements.
We present ongoing work on a set of adaptive mechanisms that benefit from the
flexibility of our design to enable querying systems to dynamically adapt to
query and write operation workloads by adjusting the configuration.

\section{Distributed Querying}
Distributed data stores achieve scalability by sorting and partitioning data
entries using their primary keys.
In such systems, queries that retrieve objects based on their secondary
attributes can be performed by executing a scan of the entire dataset, where
each object is inspected and matching objects are added to the query response.
Moreover, all data store partitions need to be scanned for matching objects, as
objects are partitioned using their primary keys, which limits the scalability
of these queries.

Instead of performing a scan, index structures can be constructed over
secondary attributes in order to efficiently identify objects that match a
given query.
Typically, a secondary index contains a set of entries, one for each existing
value of the indexed attribute.
Each entry can be seen as a key-value pair, where the key is an attribute value
and the value is a list of pointers to all objects that have this value.
The primary keys of objects are usually used as pointers.

While indexing is an efficient approach for supporting queries on secondary
attributes, maintaining indexes in distributed storage systems that
store large volumes of data incurs high storage and maintenance overheads.
In cases where queries on secondary attributes are not frequent and query
latency is not a critical requirement, the benefits of indexing may not outweigh
these costs.
In more complex scenarios, only certain attributes might be queried regularly
and require minimal query latency while others need to be searchable with
fewer requirements.
A promising approach can be a querying system that efficiently combines
secondary indexing and data store scans and performs optimisations such as
caching of search results.

\section{Modular Querying Systems}
In this section we present our main contribution, a modular design for building
geo-distributed querying systems.
We start by presenting our design and discussing how it can be used to build
distributed querying systems.
We then demonstrate the usability of our approach by applying it to two use
cases with different requirements, and finally describe a set of adaptive
mechanisms that enable querying systems to optimise their configuration.

\subsection{System Design}
A modular querying system is assembled from indexing and query processing
modules called Query Processing Units (QPUs).
QPUs operate as services that receive and process queries.
Individual QPUs perform basic indexing and query processing tasks such as
maintaining indexes, caching search results, and federating query processing
over geo-distributed datasets.
All QPUs expose a common interface that can be used to connect them to
different configurations.
Querying systems are built by interconnecting QPUs in a network.
Queries on secondary attributes are processed by being routed through the QPU
network; individual QPUs partially process a given query, forward it through the
QPU network, and combine the retrieved results.

QPUs are categorised to different classes based on their functionality:

\textbf{Data Store QPU:} Data store QPUs (dsQPUs) are connected to the
underlying data store and process queries by scanning the dataset.
Each dsQPU can be connected to a single data centre of a geo-distributed data
store.

\textbf{Index QPU:} Index QPUs (iQPUs) maintain indexes and use them to
efficiently process queries on secondary attributes.
They construct their index structures by receiving write operations from the
underlying data store.
Same as dsQPUs, each iQPU can be connected and maintain an index for a single
data centre.

\textbf{Index Merge QPU:} Index Merge QPUs (mQPUs) maintain secondary indexes by
merging updates from multiple ingest points to a global index.
They can be connected to other mQPUs, iQPUs or the underlying data store.
They can be used in use cases with disjoint datasets, such as systems that use
partial replication, in order maintain a global index for all objects in the
system.
mQPUs are responsible for resolving conflicts in index entries when merging
secondary indexes.

\textbf{Federation QPU:} Federation QPUs (fQPUs) provide the ability to query
disjoint datasets without maintaining global indexes.
A fQPU can be connected to a set of QPUs of any type;
when receiving a query, the fQPU forwards it to the QPUs it is connected to,
retrieves the corresponding responses, and combines them to a final response.
fQPUs can have more advanced functionality; for example a fQPU can forward
queries to a subset of the QPUs it is connected to based on the data placement
scheme.

\textbf{Cache QPU:} Cache QPUs (cQPUs) are used to achieve efficient query
processing by caching search results and replicating indexes to multiple
geographic locations.
A cQPU can maintain a cache storing recent query responses of any QPU, and use
it to efficiently respond to queries.
Alternatively, a cQPU can be connected to an iQPU or mQPU and act as a passive
replica by replicating index entries through a push or pull mechanism.

\begin{figure}
  \begin{center}
    \includegraphics[width=0.7\columnwidth]{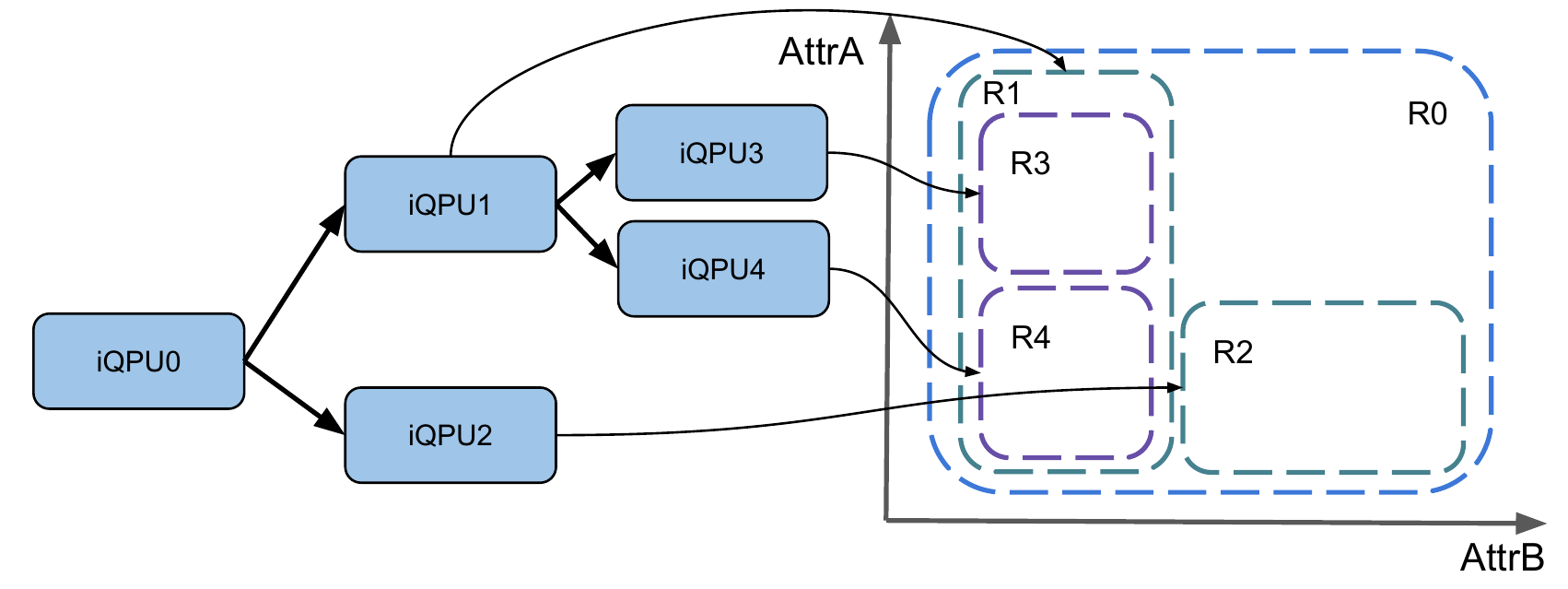}
    \caption{An example of hyperspace partitioning using an index QPU hierarchy.
    }
    \label{fig:r_tree}
  \end{center}
\end{figure}

\subsection{Secondary Index Design}
\label{index_design}
Indexing QPUs (iQPUs and mQPUs) maintain secondary indexes for a set of
secondary attributes, specified by the system configuration.
They organise their secondary indexes using hyperspace partitioning
\cite{Escriva:2012:HDS:2342356.2342360}.
Secondary attribute values are organised in a multi-dimensional space, in which
the dimensional axes correspond to secondary attributes.
Objects are mapped to coordinates on the hyperspace based on the values of
their secondary attributes.
Each QPU is responsible for responding to queries for a region of the
hyperspace, defined by value ranges [L\textsubscript{i}, U\textsubscript{i}]
for a set of secondary attributes Attr\textsubscript{1}, Attr\textsubscript{2},
..., Attr\textsubscript{N}, and maintains an index for that region.

Query operations have the same form; a query specifies a set of attributes and
the values that they must match, which can be ranges or specific values.
Based on the specified attribute values each query uniquely maps to a region
of the hyperspace.
Indexing QPUs use their indexes to respond to the query by retrieving all
objects that fall within this region.

This index design naturally supports multi-attribute and range queries and is
flexible as it allows QPUs to assign parts of their hyperspace regions to other
QPUs, and forward queries that correspond to these regions to them.
Using this mechanism, the hyperspace can be partitioned among multiple indexing
QPUs, organised in hierarchy, as shown in Figure \ref{fig:r_tree}.
The hierarchical structure of iQPUs forms a distributed index organised as an
R-tree \cite{Sellis:1987:RDI:645914.671636}.

\subsection{Index Maintenance}
When a write operation occurs, secondary indexes need to be updated by creating
index entries corresponding to the object's secondary attributes, and removing
old ones in case of an update to an existing object.
This may require distributed operations because objects and index structures may
be stored on different servers.
Moreover, creating new index entries and deleting old ones may require
contacting multiple QPUs.
Performing index maintenance operations in the critical path of updates can
therefore significantly increase write latency.

Our approach performs index maintenance in the background, in order to avoid
incurring overhead to write operations.
Special processing modules (\textit{Filter QPUs}) are responsible for
asynchronously propagating client updates from the storage system to indexing
QPUs in a streaming fashion.
If two QPUs are responsible for indexing a distinct part of the attribute value
space, updates can be propagated in parallel to them.

Propagating client updates asynchronously entails that secondary indexes are
eventually consistent with the base data;
index entries may temporarily be stale, not containing the effects of recent
write operations.
This can lead to two types of inconsistencies.
First, there can be cases in which objects are not retrieved despite matching
a query, because the corresponding index entries have not yet been created.
Second, it may be possible for a query to find entries that refer to objects
that no longer exist or have been updated with different attribute values.
Our approach permits the first type of inconsistencies, and we present
mechanisms for reducing the inconsistency window, including adjusting the amount
of resources available for index maintenance.
We avoid the second type of inconsistencies by using a mechanism that rechecks
the actual secondary attributes of each object contained in a query response and
only returns objects that match the query predicate.

\subsection{Query Processing}
Modular querying systems process queries by routing them through their QPU
networks.
As all QPUs expose a common interface, they can receive queries from users as
well as perform queries to each other through their connections.
Depending on its functionality, a QPU can respond to a given query using its
index structures, cache or by scanning the data store.
In the case of a fQPU, or if a full response cannot be obtained, the query is
decomposed to a set of sub-queries based on the characteristics of the QPU's
connections, which are then forwarded through the network.
The QPU then retrieves and combines the corresponding partial responses to
calculate the final response.

This process is performed recursively at each query processing unit.
A given query is thus incrementally decomposed to more fine-grained sub-queries
as it is routed through the QPU network.
Each sub-query is processed by a separate QPU, and partial results are then
combined to the final response.
This process requires that each QPU has local information about its
neighbours only, and not about the entire QPU network structure.

\begin{figure}
  \begin{center}
    \includegraphics[width=0.7\columnwidth]{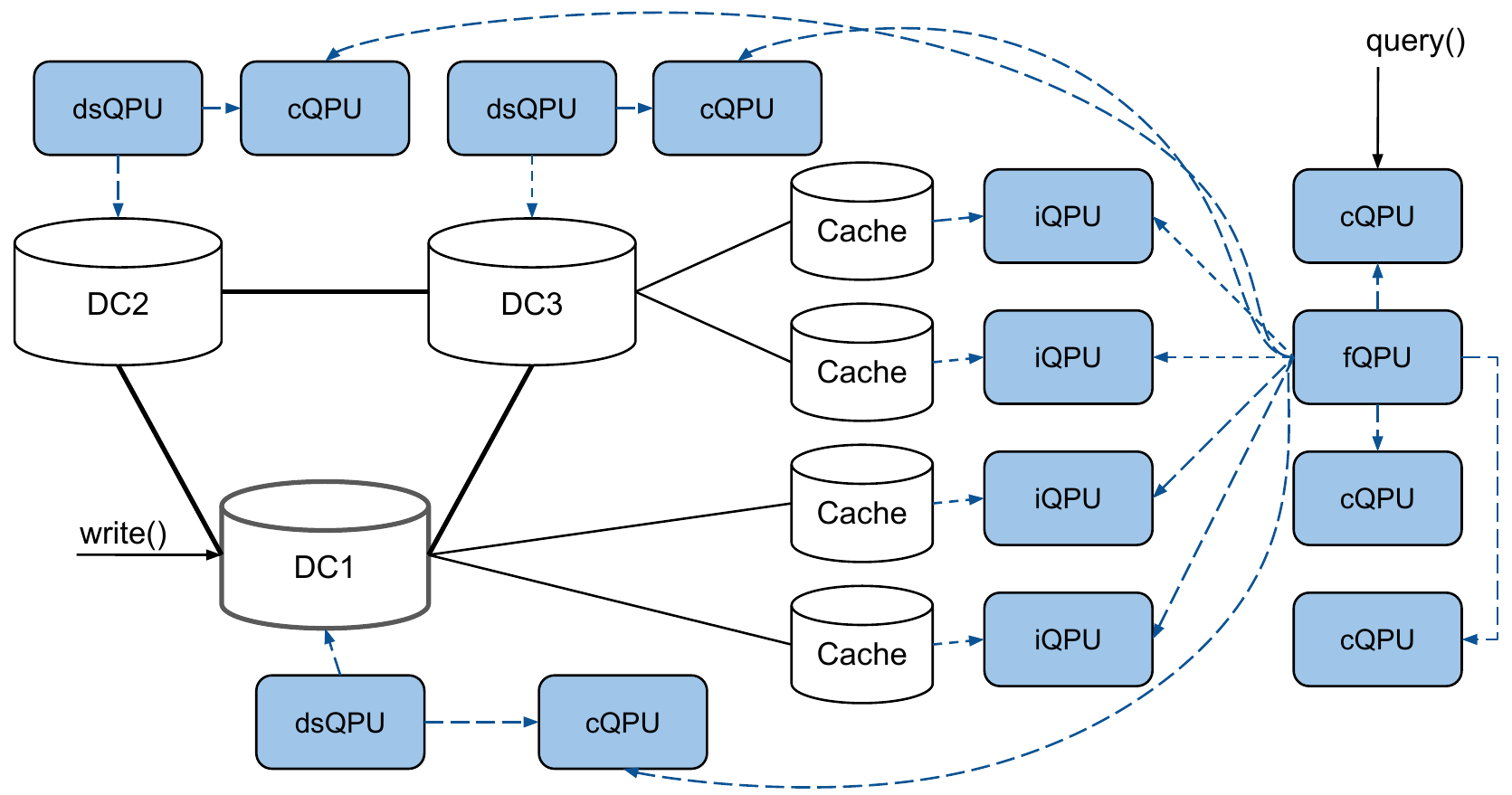}
    \caption{A QPU network configuration for querying in a content delivery
    network.}
    \label{fig:cdn}
  \end{center}
\end{figure}

\section{Example use cases}
In this section we demonstrate how the proposed modular design can be used to
enable querying in two different use cases of geo-distributed systems.

\subsection{Content Delivery Network}
Consider the example of a content delivery network (CDN) in which media files
are streamed to users distributed in multiple geographic locations.
Media files can be tagged with secondary attributes and users retrieve data by
performing queries using these attributes.
Moreover, system administrators perform queries related to data lifecycle
management operations.
The core of the system consists of a set of data centres (DCs) that replicate
every object while, at the edge of the system, cache servers store a subset of
the objects.
In this scenario, objects are inserted at the core of the system, and queries
are performed mostly from users at the edge.

A QPU network configuration for this use case is shown in Figure \ref{fig:cdn}.
A dsQPU is connected to each DC and can respond to queries by scanning the
dataset, as query latency is not critical for lifecycle queries.
A cQPU is connected to each dsQPU and maintains a cache of recent query
responses.
At the edge of the system, where applications require minimal query latency, an
iQPU is connected to each cache server to maintain a secondary index for the
objects stored there.
A fQPU is used to enable search over the entire system by forwarding queries
either to the corresponding iQPU or the core of the system, depending on the
data distribution.
Finally, cQPUs distribute recent query responses closer to the cache servers.

\begin{figure}
  \begin{center}
    \includegraphics[width=0.6\columnwidth]{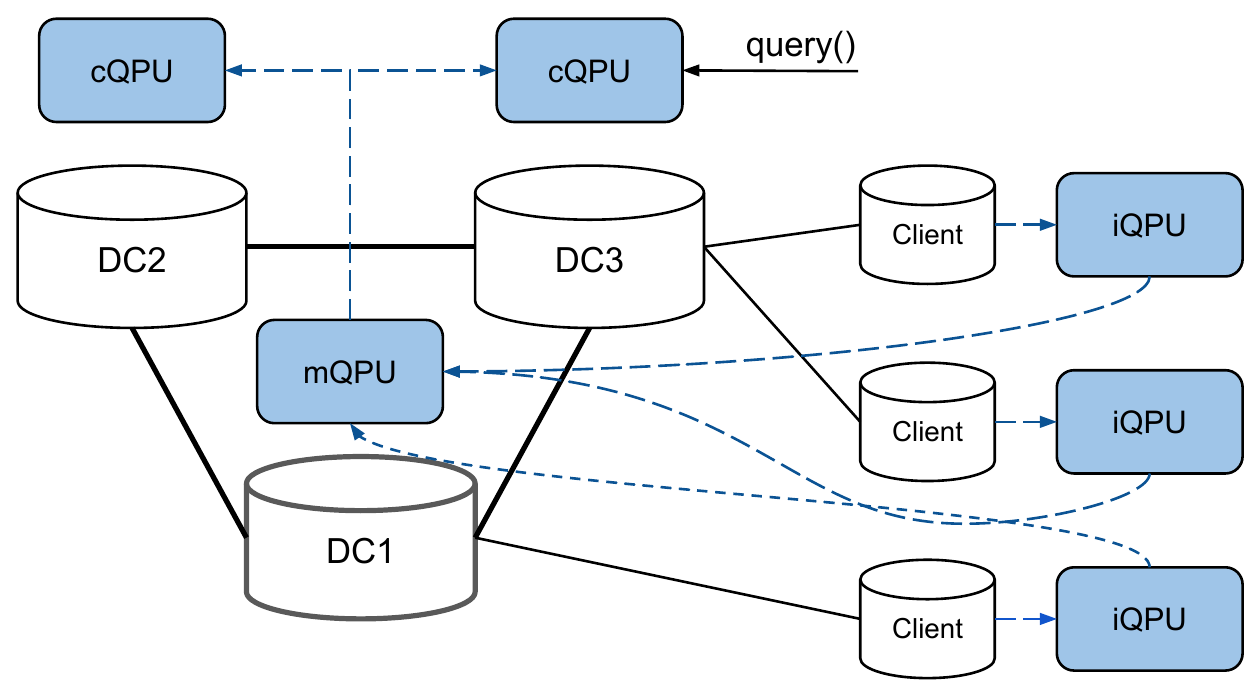}
    \caption{A QPU network configuration for querying in a data store that
    supports caching at client machines.}
    \label{fig:client_cache}
  \end{center}
\end{figure}

\subsection{Caching at client machines}
As a different use case we can consider a system that enables data caching at
client machines \cite{Preguica:2014:SFG:2709386.2709519} in order to allow
disconnected operations.
The system model is similar to the CDN use case;
at the core of the system a set of data centres replicate every object, while
at the edge each client maintains a cache storing a subset of the objects,
on which operations are performed.
Here, objects are inserted at client nodes and then propagated to the core of
the system, while queries are performed mostly at the core of the system

A QPU network that is suitable for this use case is shown in Figure
\ref{fig:client_cache}.
An iQPU is connected to each client node and maintains a local index for the
objects cached by that client.
When objects are stored, they are indexed locally at the client iQPU.
Then, a merge QPU constructs a global secondary index, and cache QPUs
replicate the global index at each DC.
In that way, the system maintains a global index, while clients store and update
local copies of parts of the global index.
Modifications to client indexes are propagated and merged to the global index.
Performing index maintenance in client machines at the edge can reduce the
computational load at the core of the system, and improve the availability of
the global index under network partition.

\subsection{Adaptive Mechanisms}
So far we have described modular querying systems as static QPU networks.
However, the flexibility of the modular design enables the implementation of
mechanisms that allow querying systems to dynamically adjust the QPU network
structure in order to adapt to query and write operation workloads.

An adaptive mechanism can be used in order to dynamically construct the indexing
QPU hierarchy described in Section \ref{index_design}.
In an initial configuration, a single QPU is responsible for maintaining an
index for the entire hyperspace for the indexed secondary attributes.
However, some regions of the hyperspace may be queried more heavily than others.
Using this mechanism, when the query load in a region of the hyperspace reaches
a threshold, a new indexing QPU is spawned and assigned with indexing and
responding to queries for that region.
Alternatively, when the query load of hyperspace region drops below a threshold,
the corresponding QPU is merged with QPUs of neighbouring regions to prevent
over-segmentation of the hyperspace.
Using this mechanism, QPU networks can dynamically adjust their structure to
balance query load among QPUs.

QPUs operate as services and are not bound to specific physical machines.
This enables the implementation of another mechanism that can dynamically adjust
the amount of computation resources available for indexing and query processing.
Heavily loaded QPUs can migrate to new physical machines in order to have more
computation resources available, while under-utilised QPUs can be collocated on
the same physical machine.

\section{Final Remarks}
Designing a geo-distributed querying system that is efficient for all use cases
is not feasible.
In this paper, we have proposed a modular architecture for building multiple
querying systems from the same building blocks, each targeting different use
case needs.
We demonstrated the usability of our approach by showing how it can be applied
to different geo-distributed system scenarios.
The design presented in this paper allows querying systems to dynamically adjust
their configuration.
This ability can be used by various mechanisms to enable querying systems to
dynamically adapt to write and query workloads.
We are currently implementing the proposed system in order to apply it to
various geo-distributed system use cases and evaluate its behaviour and
performance.

\bibliographystyle{unsrt}
\bibliography{main}

\end{document}